\documentclass{PoS}

\title{Limits on spin-dependent WIMP-proton cross-sections from the
  neutrino experiment\\ 
at the Baksan Underground Scintillator Telescope}

\ShortTitle{Baksan upper limits on SD WIMP-proton cross-sections}

\author{\speaker{Olga Suvorova}\\
        Institute for Nuclear Research of Russian Academy of Sciences,\\
prospect 60-th October 7A, Moscow 117312, Russia\\
        E-mail: \email{suvorova@cpc.inr.ac.ru}}

\author{{Musabi Boliev}\\
        Institute for Nuclear Research of Russian Academy of Sciences,\\
Baksan Neutrino Observatory, Kabardino-Balkariya 400900, Russia\\
        E-mail: \email{boliev2005@yandex.ru}}

\author{{Sergei Demidov}\\
        Institute for Nuclear Research of Russian Academy of Sciences,\\
prospect 60-th October 7A, Moscow 117312, Russia\\
        E-mail: \email{demidov@ms2.inr.ac.ru}}

\author{{Stanislav Mikheyev}\thanks{We dedicate this paper to the
    memory of our colleague S.P.Mikheyev, who died on 2011 April
    23. From the begining of operation of the Baksan Underground
    Scintillator Telescope till recently Stas was a leader of the
    neutrino experiment and he is greatly missed. His carrier spanned
    widely of neutrino investigations including famous MSW matter
    effect in a theory of neutrino oscillations.}\\ 
        Institute for Nuclear Research of Russian Academy of Sciences,\\
prospect 60-th October 7A, Moscow 117312, Russia\\ 
        E-mail: \email{mikheyev@pcbai10.inr.ruhep.ru}}

\abstract{We present updated results of the Baksan Underground
  Scintillator Telescope in search for a signal from dark
    matter particles annihilating in the center of the Sun. Based on
    the performance of the Baksan telescope measuring upward through-going 
  muons with energy larger 1 GeV, we confirm the absence of observable excess 
  in arrival directions towards the Sun relatively the expectations from
  neutrinos of atmospheric origin for 24.12 years of live time. 
  We conclude that the 90\% C.L. upper limits on elastic scattering
  cross sections of dark matter WIMP (Weakly Interactive Massive Particle) 
  off proton are comparable with correspondent 
  limits of other operating neutrino telescopes. The best
  value of the limit is $3 \times 10^{-4}$ picobarn in spin-dependent
  (SD) interaction of WIMP on proton for mass range about $100\div200$ GeV.\ }

\FullConference{VIII International Workshop on the Dark Side of the
  Universe,\\ 
  June 10-15, 2012\\
  Rio de Janeiro, Brazil}

\begin{document}

\section{Introduction}\label{sec:level1}
Neutrino channel in multiwave searches for relic dark matter (DM)
particles is accessible within data analysis of neutrino telescopes in
their regular observations of local sources like the Sun where DM
could be gravitationally trapped and further accumulated for the
time of solar system age. Up-to now there are no hints on excess of neutrino
events in the direction towards the Sun as compared with expected background
of atmospheric neutrinos at all neutrino telescopes.
However, these observations allow to set the upper limits on properties of
dark matter particles, in particular, the limit on the cross sections
of elastic scattering of WIMP on nucleons. 
Due to the Sun chemical composition with approximately $73\%$ of hydrogen, the most
sensitive quantity is the spin dependent cross sections of dark
matter WIMP on proton as it can be  seen from the latest results of
SuperKamiokande~\cite{Tanaka:2011uf} and
IceCube~\cite{IceCube:2011aj} collaborations. 

The sensitivity of neutrino telescopes strongly depends on their muon
energy threshold 
$E^{th}_{\mu}$, especially in probing of light masses of WIMP.
The Baksan Underground Scintillator Telescope~\cite{Baksan:81}
(further referred as the Baksan or the BUST) has muon energy threshold around
1~GeV, allowing for searches for neutrino signal from annihilation of
relatively light WIMPs in the Sun.
Minimal mass for WIMPs which could be probed at the Baksan is about
10~GeV. This is intrinsic challenge in light of recent claimed
spectral peculiarities and their interpretations by many authors as an
evidence on existence of light relic WIMPs both in low background
detectors of nuclear recoil reactions (see review by J.Cooley at this
conference) and in cosmic ray experiments of $\gamma$, $e^-e^+$ and
$p\bar{p}$ measurements (see review by A.Morselli at this conference). 

Here we present the preliminary results of the Baksan experiment
with the statistics twice as compared to earlier
results~\cite{Baksan:96},~\cite{Baksan:97} 
and with a new analysis improved in several ways. Firstly, the previous Baksan
results~\cite{Baksan:96},~\cite{Baksan:97} were obtained in the framework of a specific
model - MSSM (Minimal Supersymmetric Standard Model) with neutralino
as a dark matter candidate (see review of N.Fornengo). Performing the scan over its parameter
space but fixing the mass of dark matter particle, there have obtained 
conservative upper limits on muon flux and the rate of dark matter
annihilation in the Sun. Now rather then studying a specific model we follow 
present day tendency and consider specific dark matter annihilation channels,
namely annihilations into $b\bar{b}$,  ${\tau^+\tau^-}$ and
${W^+W^-}$. This was also done because the previous tactics makes it
difficult to compare the Baksan limits with the results of other
experiments. Second, we take into account current knowledge about neutrino properties in solving 
the transport of oscillating and interacting neutrinos produced in the center of the 
Sun with energy larger few GeV. 

\section{Experiment, triggers and sample of upward through-going
  muons}\label{sec:level2} 
The Baksan Underground Scintillator Telescope is 34-years continuously
operating cosmic rays detector with 4$\pi$-geometry for arrival 
penetrating charged particles. Main parameters of the telescope
have been presented in details elsewhere~\cite{Baksan:81},~\cite{Baksan:96}. Separation of
arrival directions between up and down hemispheres is made by
time-of-flight method with a time resolution 5 ns~\cite{Baksan:81}. It was
shown~\cite{Baksan:81} that 95$\%$ of $1/\beta$ values (ratios of speed  of
light and measured particle track velocity) lie in range of
0.7--1.3~ns for single downward going muons. Such interval but with
negative 
sign is used  to select upward going muons generated by neutrino
interactions in down hemisphere.  
 
Telescope is located at the altitude 1700 m above see level in the
Baksan valley of the North Caucause and at the depth 850 g/cm$^2$
under the mountain Andyrchi where the flux of atmospheric downgoing 
muons is reduced in 5000 times but it is still higher than upgoing
muons flux in six orders of magnitude. Trajectories of penetrating 
particles are reconstructed by the positions of hit tanks, which put 
together a system of 3,150 liquid scintillation counters of standard
type $(70~{\rm cm} \times 70~{\rm cm} \times 30~{\rm cm})$ in
configuration of parallelepiped $(17~{\rm m} \times 17~{\rm m} \times
11~{\rm m})$. The counters entirely cover all its sides and two
horizontal planes inside at the distances 3.6~m and 7.2~m from the
bottom. Each plane is separated from another one by concrete absorber
of $160~{\rm g}/{\rm cm}^2$. The configuration provides $1.5^{\circ}$
of angular accuracy. 

There are two hardware triggers used to select upward going
muons. They reduce initial rate of downward going muons approximately
by factor of $10^3$. Trigger I covers the zenith angle range
$95^{\circ} \div 180^{\circ}$ while trigger II selects horizontal
muons in the range $80^{\circ} \div 100^{\circ}$. The hardware trigger
efficiency of 99\% has been measured with the flux of atmospheric
muons. These two triggers give about 1,800 events per day for further
processing. In the year 2000 the telescope data acquisition system was
upgraded and allowed to simplify the trigger system down to one
general trigger with the rate of 17~Hz. All raw information is than
undergone further selection using off-line reconstruction code. 

Selection criteria of neutrino events in the Baksan neutrino
experiment did not changed since first data analyses
~\cite{Baksan:81}. Few requirements to be satisfied in
preliminary selection were determined before the telescope was
launched:  only single trajectory of penetrating particle and a
negative value of measured $\beta$. Additionally all events with
negative values of $\beta$ were scanned by eyes to check possible
misinterpretation. Also required to have an enter point of each
trajectory lower than exit point in a range not less than a tank
sizes. To reduce background from downward going atmospheric muons
coming from smaller thickness of mountain but be scattered at large
angles, all events with azimuths $0^{\circ}<\phi<180^{\circ}$ in a
sample of trigger II are rejected. It was found 1700 upward through
going muons for 24.12 years of live time (l.t.)  survived these cuts
among detected events since December of 1978 till November of 2009. 

\begin{figure}[!htb]
\begin{center}
\begin{picture}(200,150)(20,25)
\put(-100,182){\includegraphics[angle=-90,width=0.55\columnwidth]{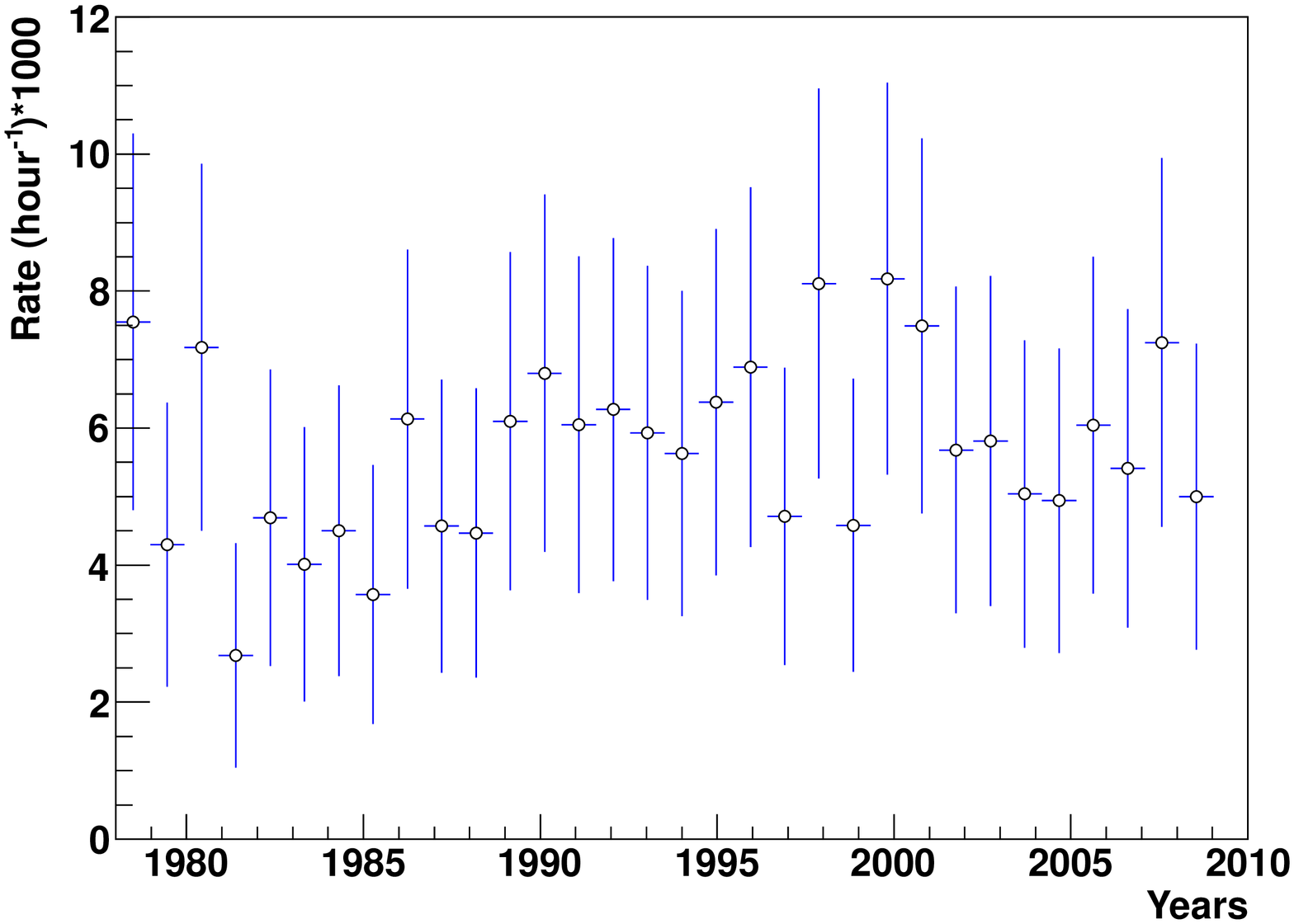} }
\put(130,20){\includegraphics[angle=0,width=0.50\columnwidth]{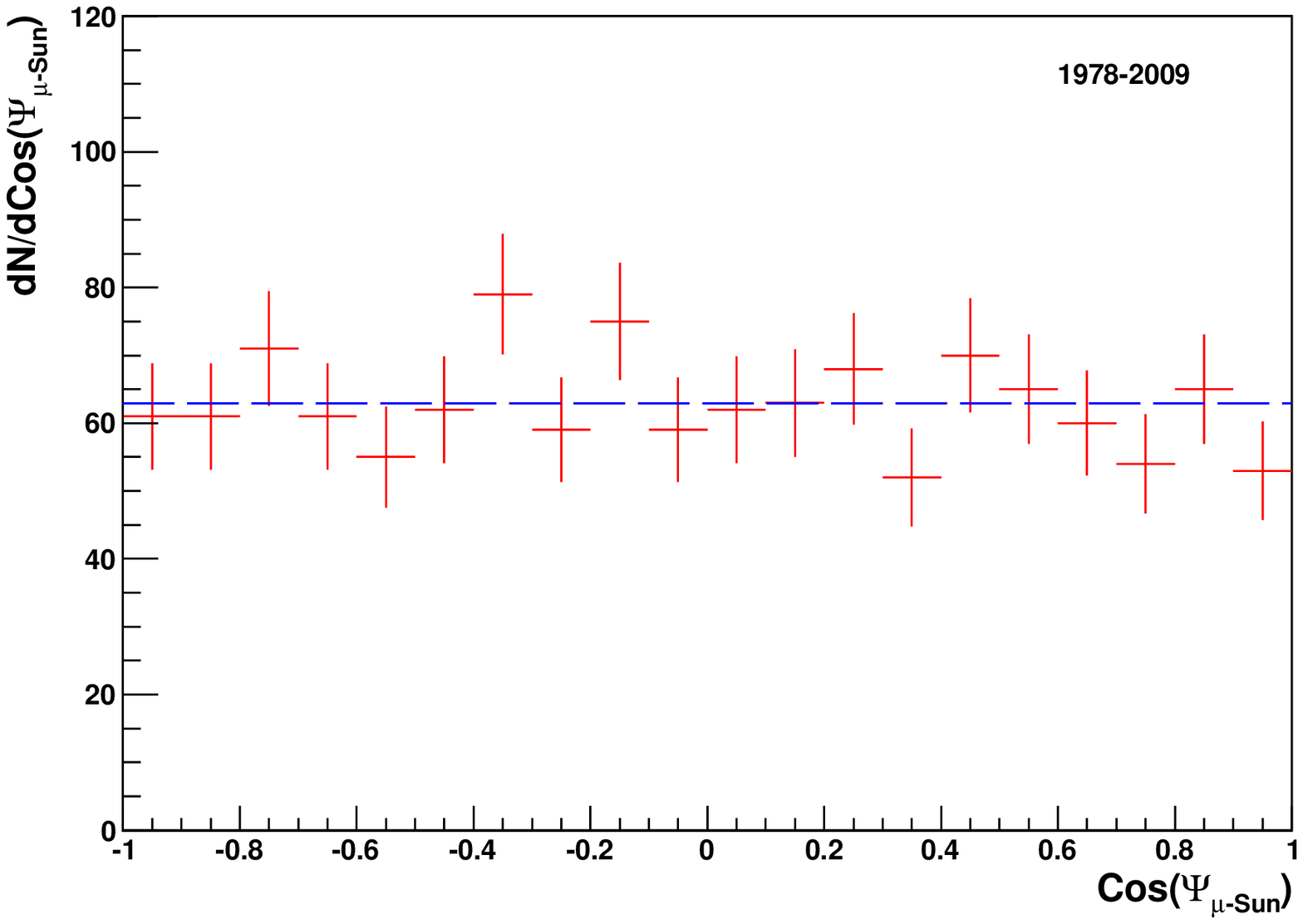}}
\end{picture}
\end{center}
\caption{\label{ris2_RateBPST-TowardSun09-20b} Left: Measured rate of
 upward through going muons at the BUST since December 1978. Right:
 Distribution of measured events in angle $\Psi_{\mu-{\rm Sun}}$  
between incoming neutrino events at the BUST and the Sun position. 
The direction of the Sun corresponds to $\cos({\Psi_{\mu-{\rm Sun}})}=1$.} 
\end{figure}

Next level of cuts impacts on events mimicked by downward going
atmospheric muon interactions or muon groups: for the reconstructed
muon trajectories its path inside the telescope had to be larger than
thickness 500 ${\rm g}/{\rm cm}^2$ (that corresponds to muon energy
threshold about 1~GeV) with both enter and exit points; selection only
events with $-1.3<1/\beta<-0.7$~ns; edge conditions on enter and exit
for events in sample of horizontal events. Totally 1255 events for
24.12 l.t. years survived all cuts. From
Fig.~\ref{ris2_RateBPST-TowardSun09-20b} (left)
one can conclude about stable rate of measured upward through-going
muons during all years of observation.

\section{The Sun survey during three decades}\label{sec:level3}
In search for neutrinos from the dark matter processes in the Sun,
we analyse distribution of measured events as a function of
$\cos({\Psi_{\mu-{\rm Sun}})}$ where $\Psi_{\mu-{\rm Sun}}$ is the angle
between upward incoming muons and the Sun position. In
Fig.~\ref{ris2_RateBPST-TowardSun09-20b} (right) the 
obtained cosine distribution is presented. The mean rate per bin is
shown by blue line. We estimate the background expected from
atmospheric neutrinos directly from real data and shifted (false) Sun  
positions. Here we follow our previous
analysis~\cite{Baksan:96},~\cite{Baksan:97} where it was shown that
this method is compatible with Monte Carlo (MC) simulations of the
detector response. Mean energy of simulated neutrinos, which produce
muons with energy larger 1~GeV to cross the 
Baksan telescope, is about 50~GeV. From comparison of data sample for 
21.15 years of l.t.~\cite{Baksan:06} and MC statistics of time larger
than real data taking by factor 22 (i.e., in total of 460 years) it
was found that the ratio of observed total number of events to
expected one without neutrino oscillations is $0.87 \pm 0.03(stat.)
\pm 0.05(syst.) \pm 0.15(theor.)$. The details has been presented
elsewhere~\cite{Baksan:06}. 

\begin{figure}[!htb]
\begin{center}
\begin{tabular}{cc}
\includegraphics[angle=0,width=0.50\columnwidth]{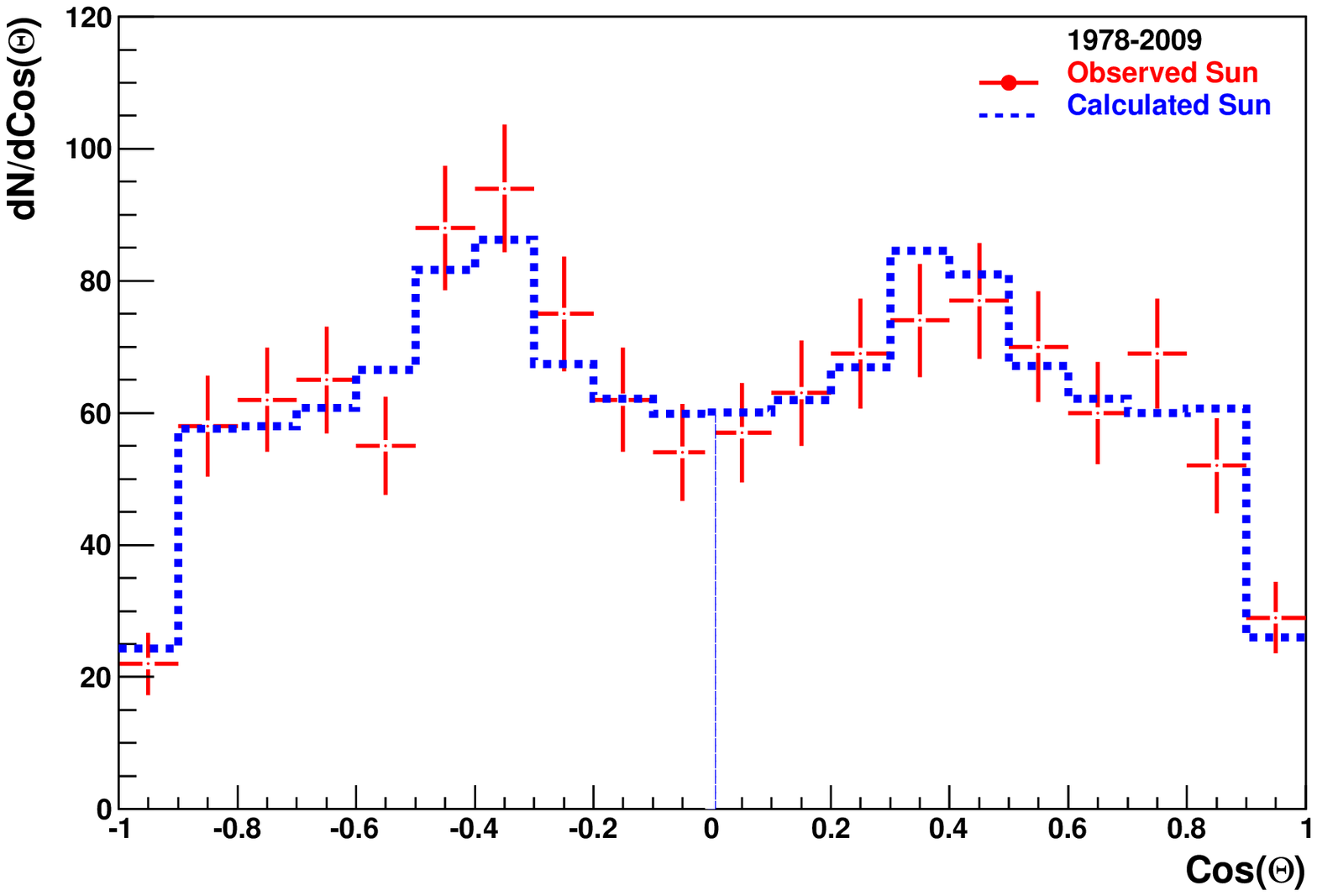} 
&
\includegraphics[angle=0,width=0.47\columnwidth]{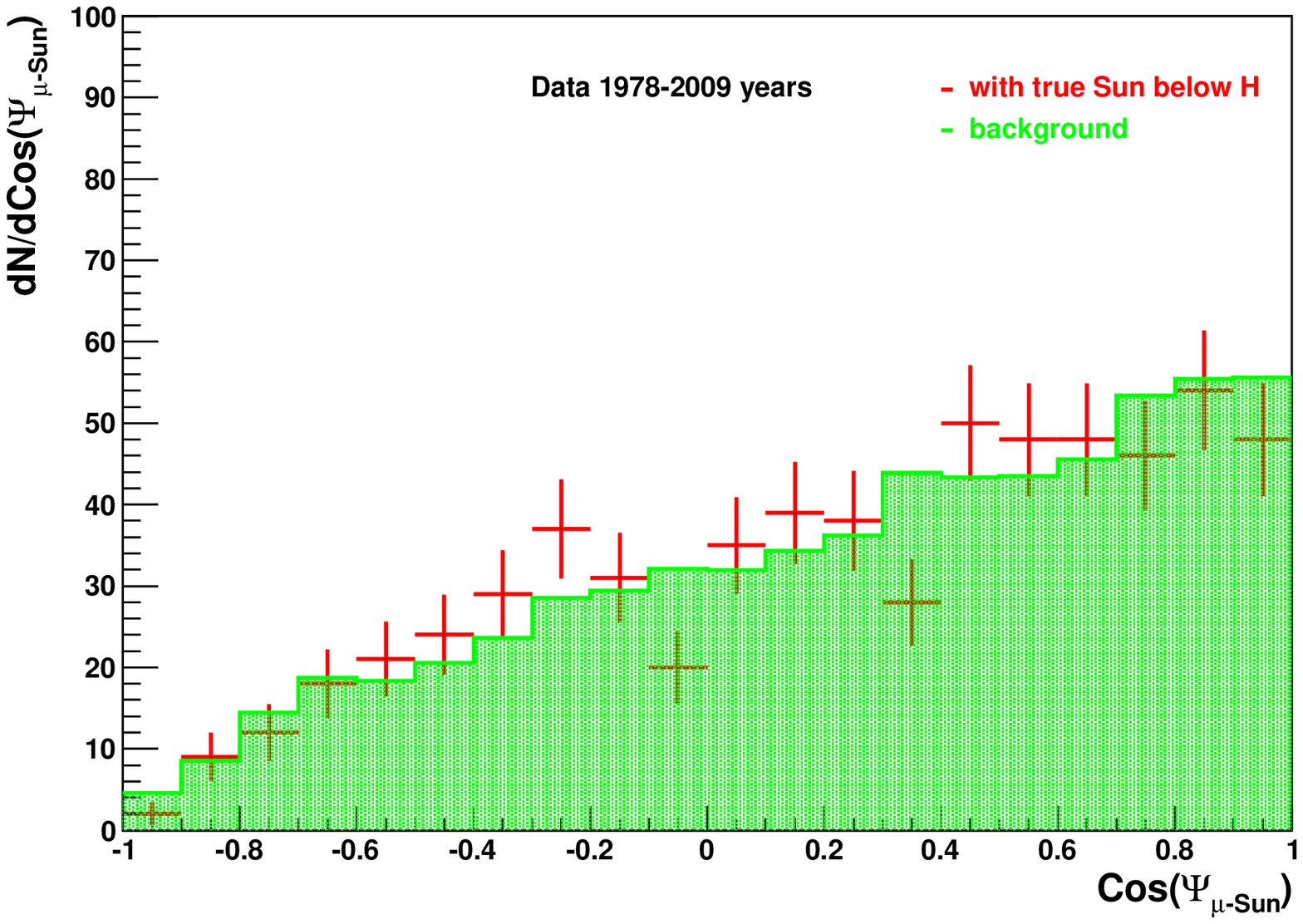} 
\end{tabular}
\end{center}
\caption{\label{SunTrack-ToFalseSun_Dw} Left: Zenith distributions of
  the Sun during the BUST runtime: observed (red 
  points with error bars) and calculated (blue, see text). Right:
  Distributions of $\cos({\Psi_{\mu-{\rm Sun}})}$ for events toward true
  Sun position below horizon (red) in comparison with measured
  background (green) obtained from the distributions for
  $\cos({\Psi_{\mu-{\rm Sun}})}$ averaged over six shifted Sun
  positions.} 
\end{figure}

At the location of the Baksan telescope (43,16$^\circ$N and
42,41$^\circ$E) the Sun is seen in average about half of time per day
during a year in both hemispheres; the real bin contents of solar
zenith ($\theta$) distributions reconstructed from data time
information (red points with error bars) and one (blue histogram)
calculated using the Positional Astronomy Library~\cite{slalib} are
shown in Fig.~\ref{SunTrack-ToFalseSun_Dw} (left). 
The reconstructed values are obtained from a
sample of 1255 selected upward going muons for 24.12 years of live
time. The calculated distribution is normalized to this number. As it
can be seen that our multi-years measurements reproduce the Sun
full-year passing track with a good accuracy. That is the base for our
further analysis. 

\begin{figure}[!htb]
\begin{center}
\includegraphics[angle=0,width=0.60\columnwidth]{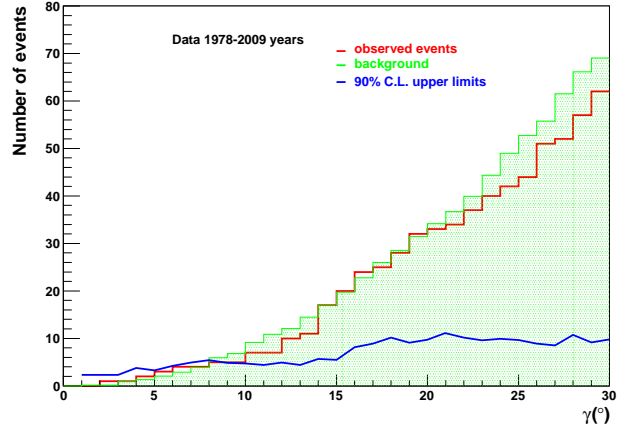} 
\end{center}
\caption{\label{AngLimToSun30-tot2009} The 90\% C.L. upper limits on
  additional number of events toward the Sun (blue) in cone half-angles $\gamma$.
  Also shown are measured events (red) and
  background (green) in the same cones.} 
\end{figure}

Potentially the neutrinos from dark matter annihilations in the Sun
could be found at night time when the Sun is below horizon. For
"night" neutrinos we compare cosine distribution in
Fig.~\ref{SunTrack-ToFalseSun_Dw} (right) and integrated angular
distribution in Fig.~\ref{AngLimToSun30-tot2009} 
with respect to measured
background which has been obtained from averaged distribution of six
cases of false Sun positions shifted over ecliptic. In Fig.~\ref{AngLimToSun30-tot2009} 
shown on abscissa value $\gamma$ is a half open angle of the cone 
toward the Sun. We observe no an excess of events coming from the Sun 
and set the 90\% C.L. upper limits on additional events in each cone
as also shown  in Fig.~\ref{AngLimToSun30-tot2009}. Assuming Poisson
statistics for both expected background ($N_B$) and observed events
($N_{obs}$) we obtain upper limits C.L. on additional number of signal
events ($N_{S}$) as follows~\cite{PDG}:
\begin{eqnarray}
{C.L.} = 1 - \frac { e^{-(N_B+N_S)} 
\sum^{N_{obs}}_0{{(N_B+N_S)^n}\over{n!}} } {e^{-N_B}
\sum^{N_{obs}}_0{{{N_B}^n}\over{n!}}} 
\label{eq:five}.
\end{eqnarray}
Further we translate these results into limits on annihilation
rates and muon fluxes from the Sun direction, considering in
details a transport of high energy neutrinos from the Sun to the
Baksan detector and found dependence on WIMP mass of value $\gamma$ where
90\% of $N_{S}$ events are collected. 
 
\section{Transport of solar neutrinos with energy higher few
  GeV}\label{sec:level4} 
In this Section we briefly describe techniques which we used for
numerical simulations of muon signal in the BUST and leave the details for
future publication~\cite{future}. Resulting neutrino spectrum from
WIMP annihilations is a mixture of inclusive spectra from decays of
produced copious particles. Spectral characteristics of this spectrum
depends on the WIMP mass and its annihilation channel. To obtain
model-independent bounds we consider WIMP of masses from 10~GeV to
1~TeV with three dominant annihilation channels: $b\bar{b}$,
$W^{+}W^{-}$ and $\tau^{-}\tau^{+}$. The channel $b\bar{b}$
represents an example of ``soft'' spectrum while $W^{+}W^{-}$ and 
$\tau^{+}\tau^{-}$ are examples of ``hard'' spectra. 

We use our own C code for MC simulation of neutrino propagation  from
the center of the Sun to the level of the Baksan detector. At the point of
production in the Sun, we take neutrino spectra for each annihilation
channel obtained in Ref.~\cite{Cirelli:2005gh}. When simulating
neutrino propagation we take into account oscillations in the Sun
matter and in the vacuum, CC (charged current) and NC (neutral
current) scattering of neutrino with interior of the Sun including
$\tau$ regeneration. Our procedure of MC simulations is very similar
to that of in WimpSIM package~\cite{Blennow:2007tw,wimpsim}. Neutrino
oscillations in $3\times 3$ scheme were implemented according to the
algorithm presented in Refs.~\cite{Ohlsson:1999xb,Ohlsson:2001et}
which is very convenient for using with varying density of the
Sun. For parameters of neutrino mixing matrix we use the best fit of
experimental data of Ref.~\cite{Tortola:2012te}. Below we present the
results for normal hierarchy only. We use standard solar
model presented in Ref.~\cite{Bahcall:2004pz}. Neutrino-nucleon DIS cross
sections and distributions are calculated according to formulas
presented in~\cite{Paschos:2001np} and CTEQ6~\cite{Pumplin:2002vw}
parton distribution functions. Finally, we calculate flux of muons
produced by neutrinos in the rock below the Baksan telescope
as in Ref.~\cite{Erkoca:2009by}. We check results obtained by our code 
for neutrino flux at the level of the Earth with that of obtained by
WimpSIM~\cite{Blennow:2007tw,wimpsim} and find an agreement with a
sufficient accuracy~\cite{future}.

\section{Results and limits on neutralino-proton cross
  sections}\label{sec:level5} 
If the capture and annihilation of WIMPs in the Sun reach exact
equilibrium for time much less than age of the solar system 4,5~Giga
years, the annihilation rate $\Gamma_{A}$ is equal to a half of
capture rate defined by WIMPs scattering cross sections off solar
matter. And in this case, the limit on muon flux from dark matter
annihilation in the Sun and the limit on annihilation rate can be
recalculated to limit on elastic cross section of dark matter particle
on nucleons~\cite{Edsjo:09,Demidov:2010rq}. According to that
annihilation rate can be divided into two pieces 
\begin{eqnarray}
{\Gamma_{A}} = {\Gamma_{A}(\sigma_{SI})+\Gamma_{A}(\sigma_{SD})} 
\label{eq:four},
\end{eqnarray}
where $\Gamma_{A}(\sigma_{SI})$ and $\Gamma_{A}(\sigma_{SD})$ are
the parts of equilibrium annihilation rate determined by either
spin-dependent (SD) or spin-independent (SI) interactions. 

\begin{figure}[!htb]
\begin{center}
\begin{tabular}{cc}
\includegraphics[angle=-90,width=0.50\columnwidth]{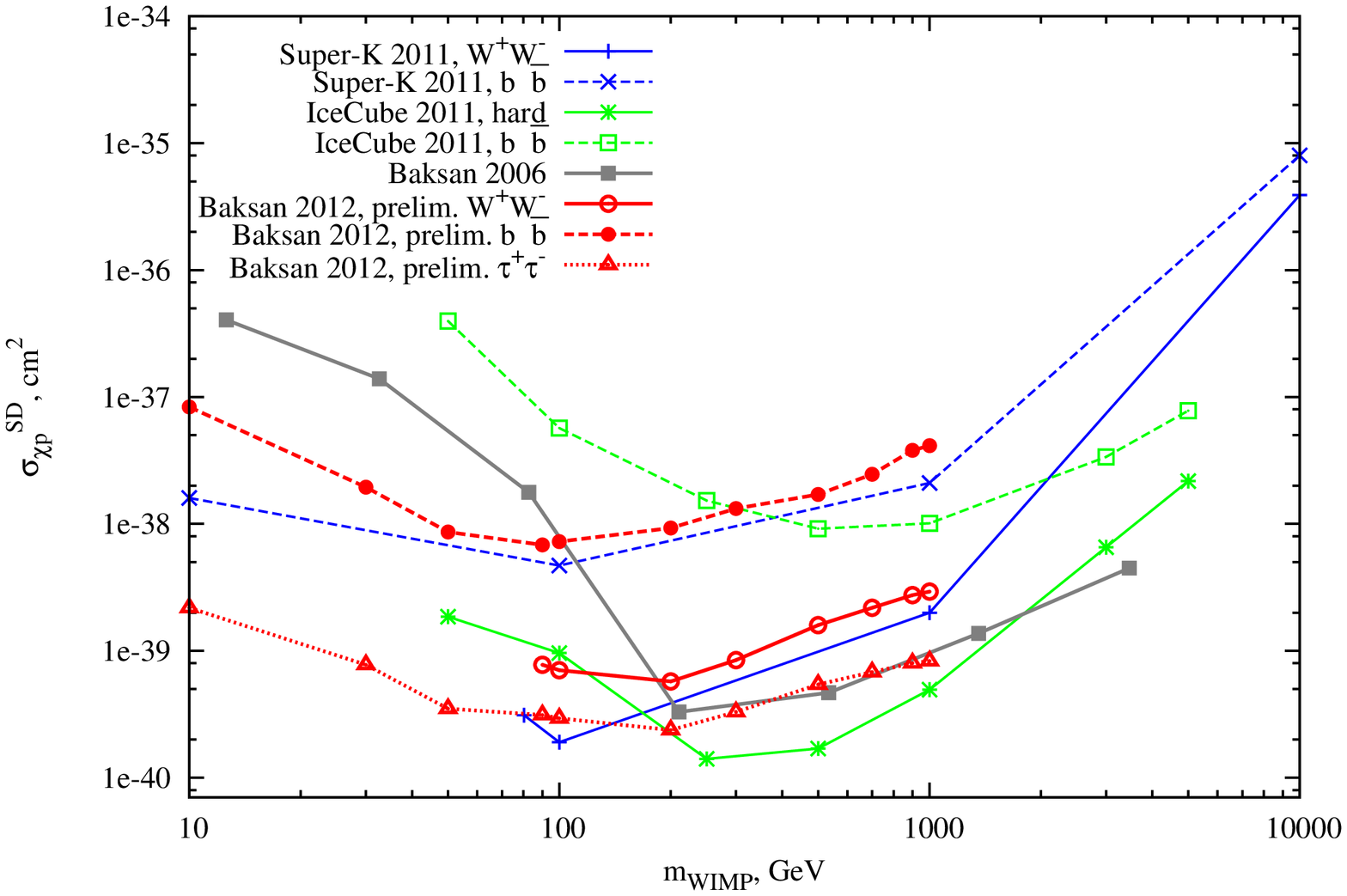} 
&
\includegraphics[angle=-90,width=0.50\columnwidth]{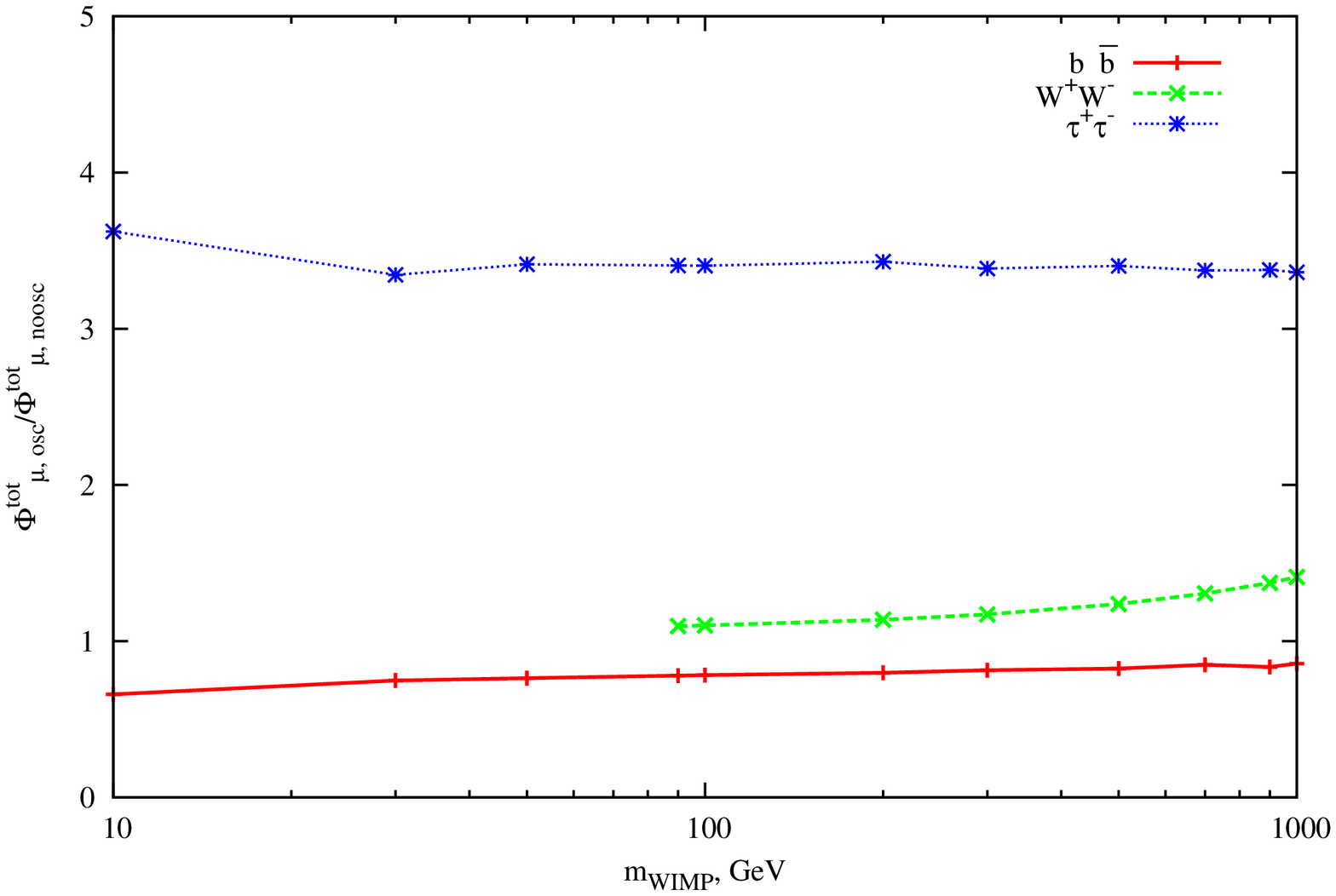} 
\end{tabular}
\end{center}
\caption{\label{sd_lim} The Baksan limits on SD elastic cross section
  of WIMP on proton in comparison with other experimental results
  (left);  ratios of expected muon fluxes both from DM neutrino 
and antineutrino generated in pure three annihilation branches
$b\bar{b}$ quarks, or ${\tau^+\tau^-}$ leptons or ${W^+W^-}$ bosons in
cases with and without three flavours oscillations (right). 
}
\end{figure}

To obtain the upper limit on SD elastic cross sections we use
recalculation procedure described in~\cite{Demidov:2010rq} where we
obtained corresponding coefficients in the following expressions
\begin{eqnarray}
\sigma^{UppLim}_{SD}(m_{\chi})= \lambda^{SD}\left( m_{\chi}\right) \cdot
\Gamma^{UppLim}_A(m_{\chi})\\ 
\end{eqnarray}
as functions of neutralino mass $m_{\chi}$. The (preliminary) results
for the limits on SD elastic cross section from Baksan's data for the
annihilation channels $W^{+}W^{-}$, $b\bar{b}$ and $\tau^{+}\tau^{-}$
are presented in Figure~\ref{sd_lim} (left)
along with corresponding limits of
SuperKamiokande~\cite{Tanaka:2011uf} and
IceCube~\cite{IceCube:2011aj}. Typically, only $W^{+}W^{-}$ and
$b\bar{b}$ channels are considered in the limit on dark matter
annihilation rate or SD elastic cross section. We include also
$\tau^{+}\tau^{-}$ channel because of two reasons. Firstly, it allows to
extend the limits on SD cross section for ``hard'' type of neutrino
spectra to lower masses of dark matter. Second, it illustrates an
important point: in spite of the fact that the limits on muon fluxes
are approximately the same for  $W^{+}W^{-}$ and $\tau^{+}\tau^{-}$,
the limits on SD cross section (and annihilation rate) are quite
different for these two channels as it can be seen in
Fig.~\ref{sd_lim} (left). 
The explanation of this fact resides partly
in the effect of neutrino oscillations: the enhancement due to this
effect for $\tau^{+}\tau^{-}$ channel is considerably larger than  
for the case of $W^{+}W^{-}$ channel. This is illustrated in
Figure~\ref{sd_lim} (right) where the ratio of muon fluxes with and
without oscillation for different annihilation channels depending on
mass of dark matter is shown. Another important effect is a breaking
"democracy" in average number of high energy neutrinos per one act of
dark matter annihilations: for $\tau^{+}\tau^{-}$ pairs it is larger
in a factor of $2\div3$ than for $W^{+}W^{+}$ or $b\bar{b}$ channels. 

It can be seen in Fig.~\ref{sd_lim} that neutrino oscillations result
in decrease of muon flux from $b\bar{b}$ annihilation channel,
which is expected to be dominating at lighter neutralino masses. At the
same time the muon flux from $W^{+}W^{-}$ annihilations is enhanced.
These results allow also to interpret more correctly the previous
Baksan limit~\cite{Baksan:97} (brown line in Fig.~\ref{sd_lim}, left)
calculated without taking into account neutrino oscillations.

\section{Summary and conclusions}\label{sec:level6}

We have performed updated analysis of indirect dark matter search
with the Baksan Underground Scintillator Telescope data for 24.12
years of 
live time. Search for an excess of upward going muons in the direction
toward the Sun we do not observe any significant deviation from
expected atmospheric background. We have presented preliminary the
90\% C.L. upper limits on spin-dependent elastic cross section of dark
matter on proton assuming particular annihilation channels $b\bar{b}$,
$W^{+}W^{-}$ and $\tau^{+}\tau^{-}$. These limits have been derived
from the limits on muon fluxes and annihilation rates for dark matter
masses in the interval $10\div1000$~GeV. The best value of the limit is
about $3 \times 10^{-4}$~picobarn for WIMP masses within
$100\div200$~GeV, that is comparable with those presented by 
SuperKamiokande and IceCube collaborations.

\begin{acknowledgments}
We acknowledge our colleagues from the Baksan Observatory for permanent 
collaboration providing a long-term stability of the telescope readout
systems. The work was supported in part by the Russian Found for Basic
Research Grants 09-02-00163a and 11-02-01528a. The work of S.D. was
supported by the grants of the President of the Russian Federation
NS-5590.2012.2, MK-2757.2012.2, by Russian Foundation for Basic
Research grants 11-02-01528-a and 12-02-31726-mol-a and by the
Ministry of Science and Education under contract No.~8412. The
numerical part of the work was performed on Calculational Cluster of
the Theory Division of INR RAS. 
\end{acknowledgments}

\end{document}